\documentclass{lncse}
\usepackage[]{graphicx}

\begin{document}

\title{Statistical Analysis of Freeway Traffic}

\author{L. Neubert\inst{1} \and L. Santen\inst{2} \and
  A. Schadschneider\inst{2} \and M. Schreckenberg\inst{1}}

\institute{Physik von Transport und Verkehr,
  Universit\"at Duisburg, Germany \and Institut f\"ur
  Theoretische Physik, Universit\"at zu K\"oln, Germany}

\maketitle

\begin{abstract}
  Single-vehicle data of freeway traffic as well as selected {\em
    Floating-Car} ({\em FC}) data are analyzed in great detail.
  Traffic states are distinguished by means of aggregated data. We
  propose a method for a quantitative classification of these states.
  The data of individual vehicles allows for insights into the
  interaction of vehicles. The time-headway distribution reveals a
  characteristic structure dominated by peaks and controlled by the
  underlying traffic states. The tendency to reach a pleasant level of
  ``driving comfort'' gives rise to new $v(\Delta x)$-diagrams, known
  as {\em Optimal-Velocity} ({\em OV}) curves. The insights found at
  locally fixed detectors can be confirmed by {\em FC} data. In
  fundamental diagrams derived from local measurements interesting
  high-flow states can be observed.
\end{abstract}

\section{Introduction}

Experimental and theoretical investigations of traffic flow have been
the focus of extensive research interest during the past decades.
Various theoretical concepts have been developed and numerous
empirical observations have been reported (see e.g.
\cite{tgf,selected_books,May,Bovy,Kerner,fvu,s2s,asrev,Koshi83,Bando94,Barlov,KK,Treiber,Wagner,Neubert}
and references therein). However, most of them are still under debate.
Recent experimental observations suggest the existence of three
qualitatively different phases \cite{Kerner}: 1) {\em Free-flow
  states} with a large mean velocity, 2) {\em synchronized states},
where the mean velocity is considerably reduced compared to the
free-flow states, but all cars are moving, and 3) {\em stop-and-go
  states} with small jams.  The last two states are summarized as {\em
  congested states}.  Following \cite{Kerner} three different types of
synchronized traffic exist.  Long time intervals of constant density,
flow, and velocity characterize the first type~(i). In the second
type~(ii) variations of density and flow are observable, but the
velocity of the cars is almost constant. Finally, there is no
functional dependence between density and flow in the third type~(iii)
of
synchronized traffic.\\
This work focuses basically on two points. First, we present a {\em
  direct analysis of single-vehicle} data which leads to a more
detailed characterization of the different microscopic states of
traffic flow, and second we use standard techniques of time-series
analysis, i.e. utilization of autocorrelation and crosscorrelations,
in order to establish {\em objective} criteria for an identification
of the different states \cite{Neubert}.  These investigations have
been carried out using extended data sets from locally fixed detectors
on a German highway. Concluding, the insights we got are
compared with results drawn from car-following experiments.\\

\section{Data from Locally Fixed Detectors}

\begin{figure}
  \begin{center}
    \includegraphics[width=.7\textwidth]{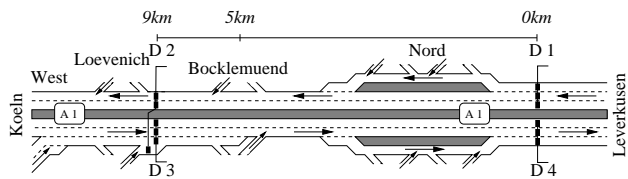}
    \caption{The examined section of the highway A1 near K\"oln. The
      detector installations D1/D4 are nearly one kilometer apart from
      a highway intersection.}
    \label{fig:roadsketch}
    \end{center}
\end{figure}

\begin{figure}
  \begin{center}
    \includegraphics[width=.38\textwidth]{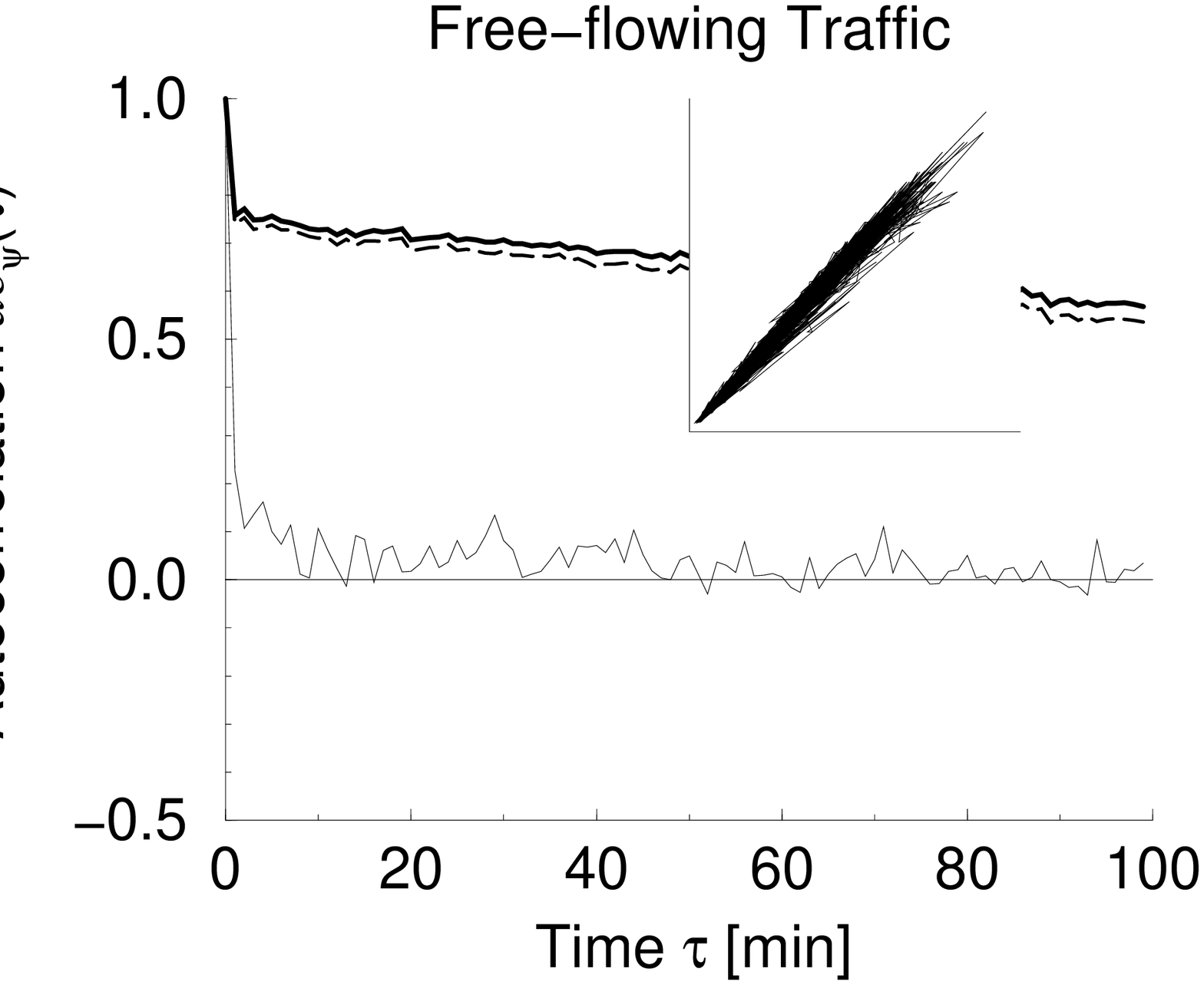}
    \includegraphics[width=.38\textwidth]{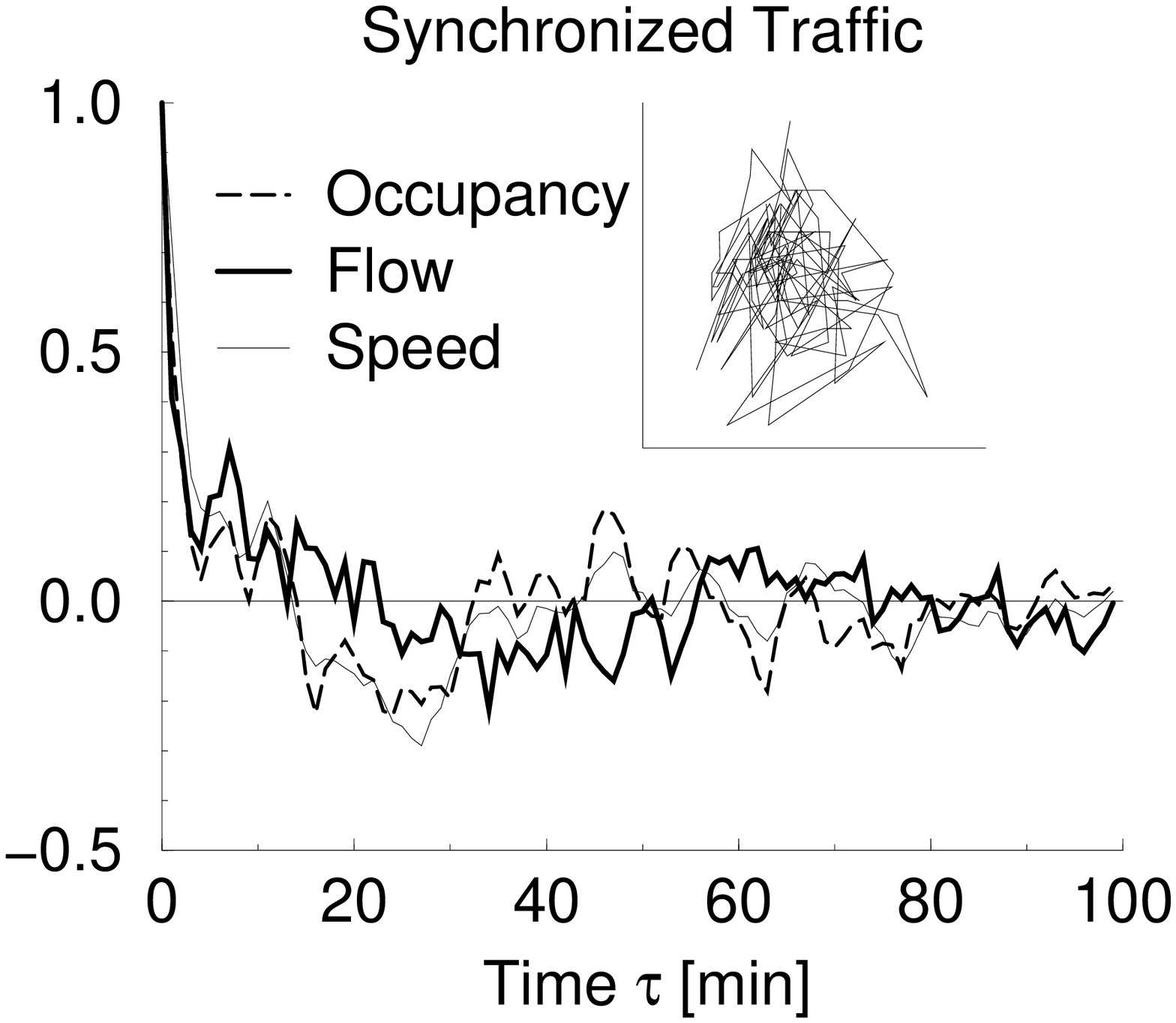}
    \includegraphics[width=.38\textwidth]{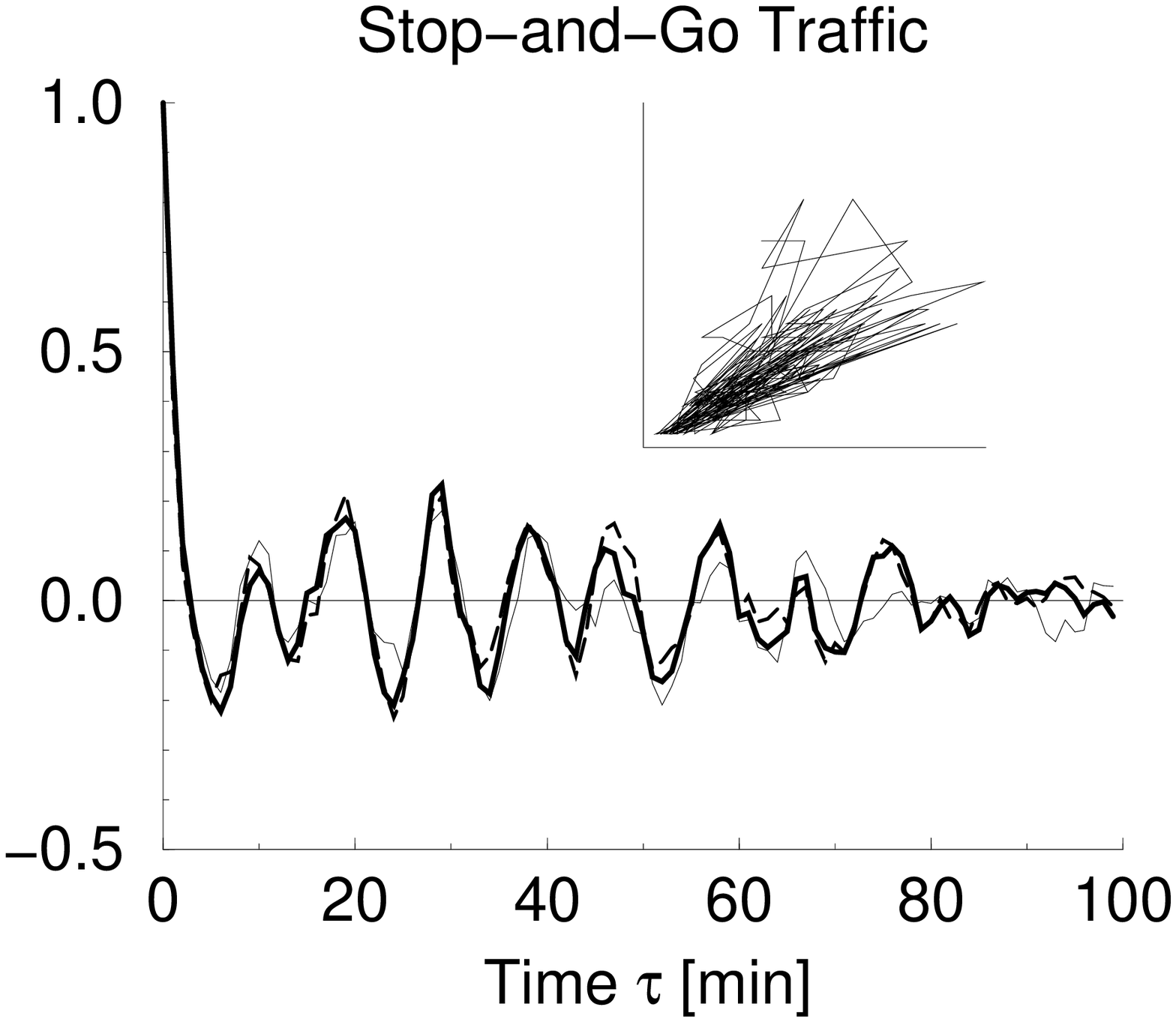}
    \caption{Typical autocorrelation functions of
    aggregated variables depending on the traffic state. The
    correlation collapses in the case of synchronized flow, in
    stop-and-go traffic one observes oscillations \cite{Kuehne}.}
    \label{fig:autocorr_states}
  \end{center}
\end{figure}

\begin{figure}
  \begin{center}
    \includegraphics[width=.38\textwidth]{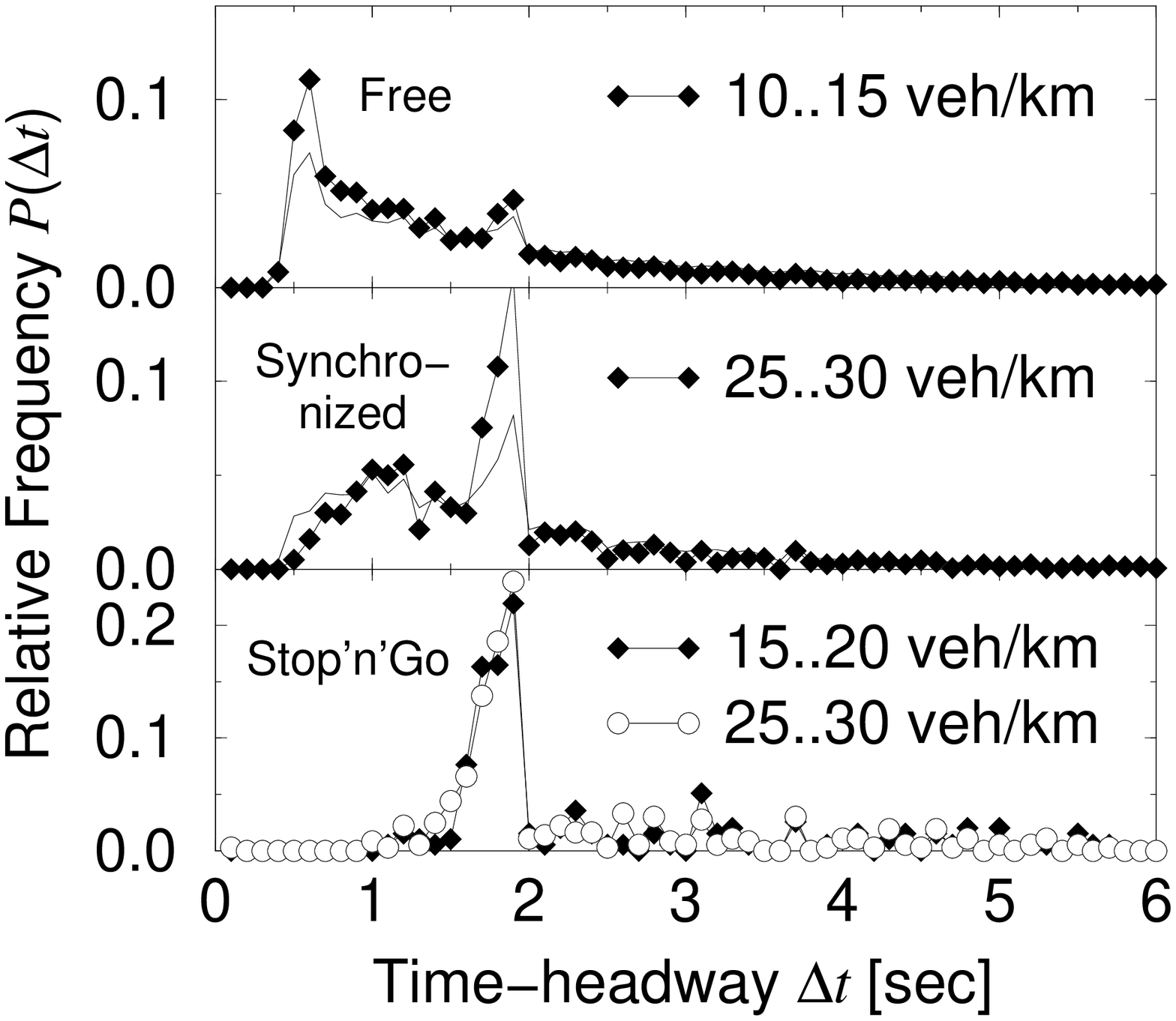}
    \includegraphics[width=.38\textwidth]{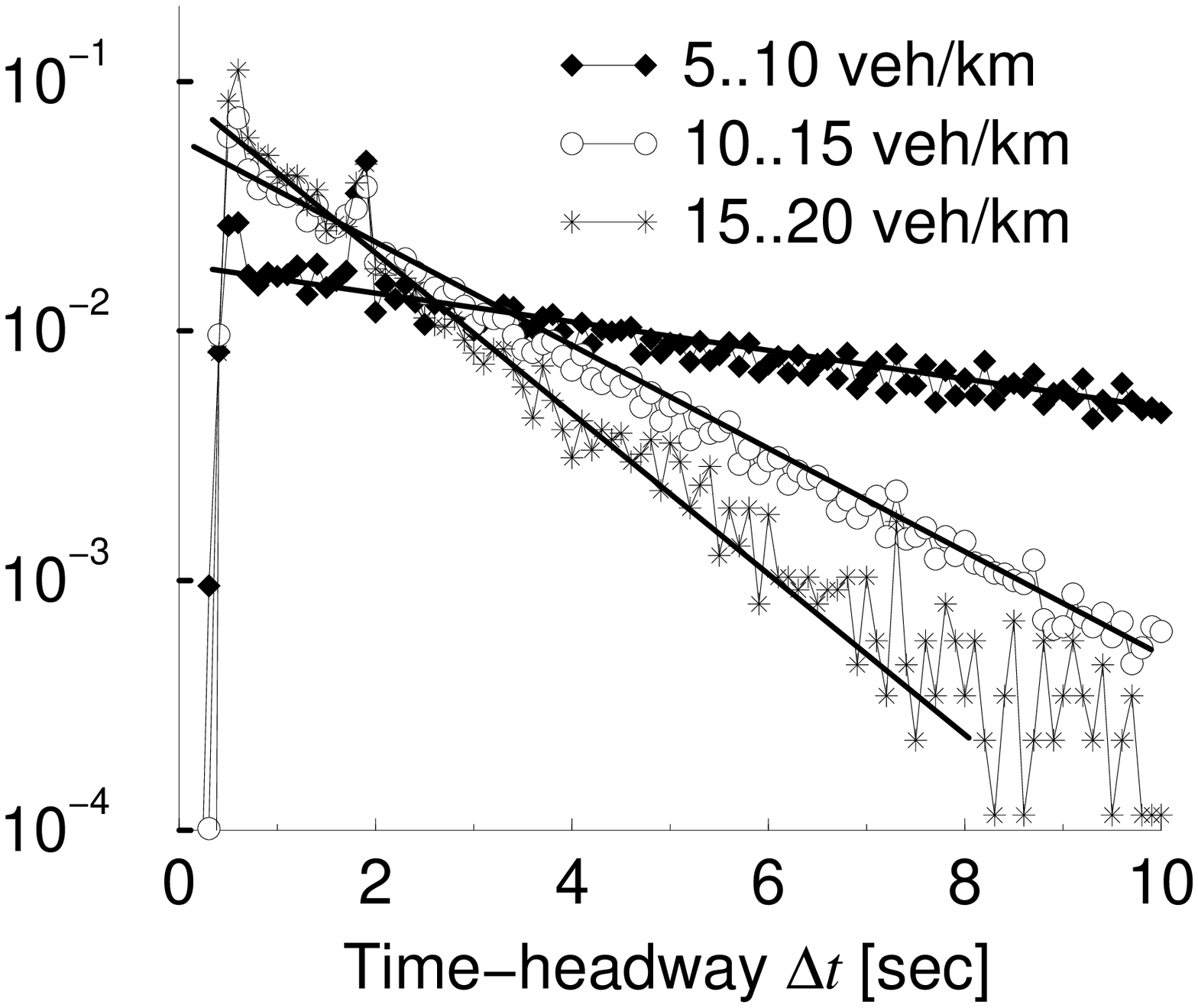}
    \caption{At least one peak emerges in all traffic states, which
    position is independent of the density of traffic. In free-flow
    traffic the background signal can be described through a Poisson
    process with exponential decay \cite{May}.}
    \label{fig:timeheadways}
  \end{center}
\end{figure}

\begin{figure}
  \begin{center}
    \includegraphics[width=.38\textwidth]{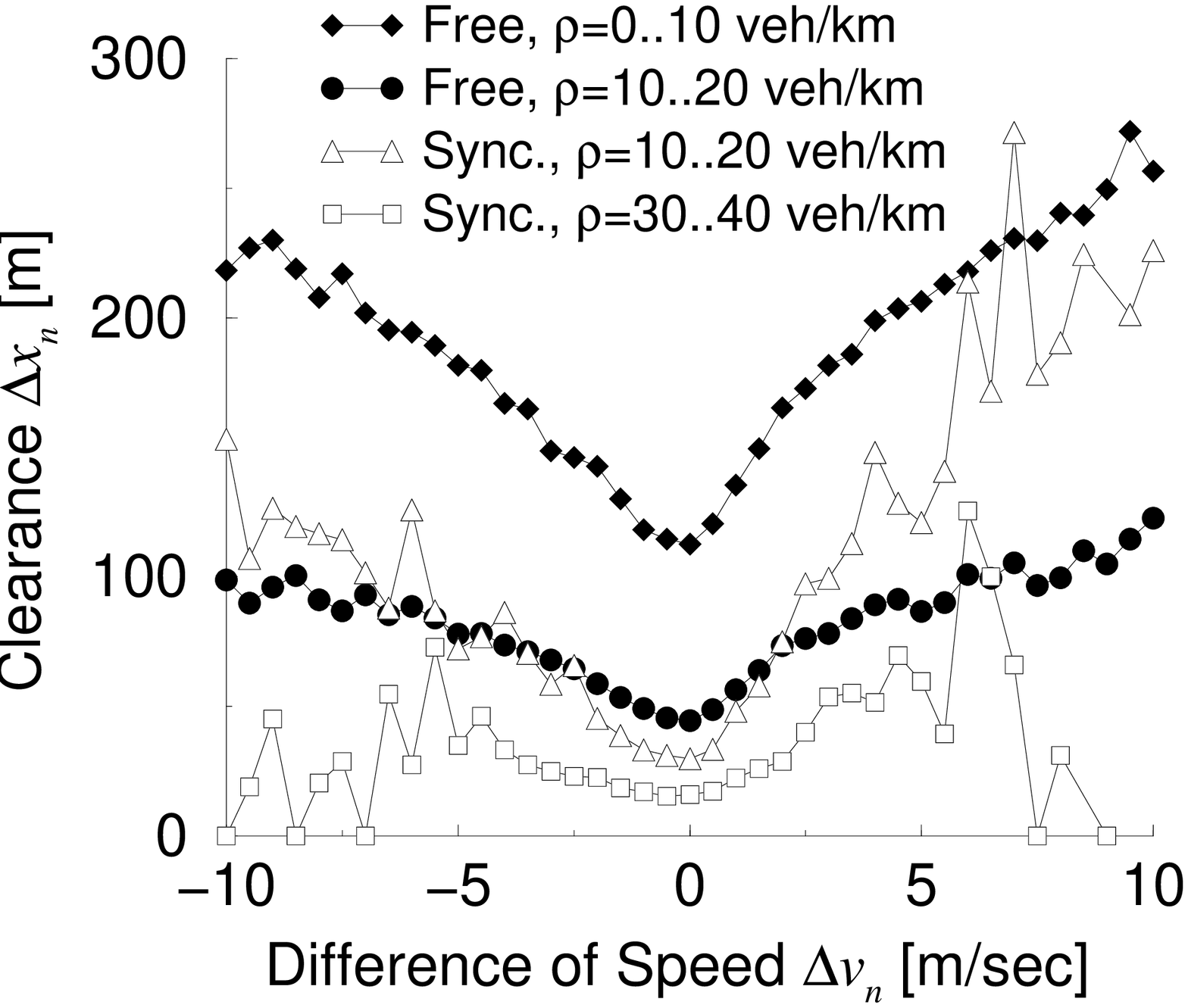}
    \includegraphics[width=.38\textwidth]{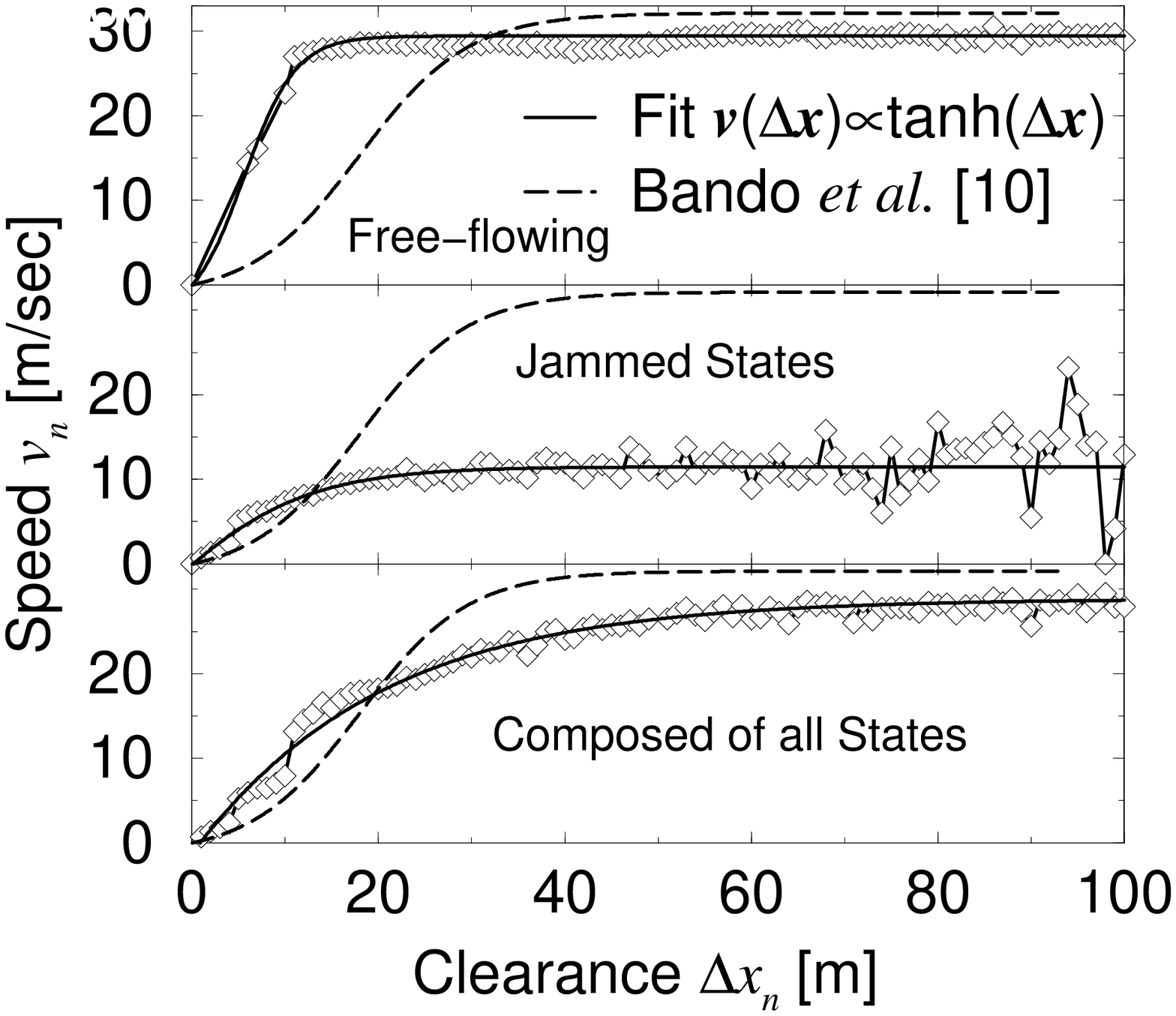}
    \caption{Left: ``Driving comfort'' is characterized by a
      minimization of the difference of speeds between two consecutive
      cars with a minimal distance. Consistently, only data points
      with $\Delta v_n\approx 0\,\mbox{m/sec}$ should contribute to
      the {\em OV} curve \cite{Bando94}.}
    \label{fig:ov}
  \end{center}
\end{figure}

The data set is provided by 12 counting loops all located at the
German highway A1 near K\"oln (Fig.\,\ref{fig:roadsketch})
\cite{Wagner,Neubert}. At this section of the highway a speed limit of
$100\,\mbox{km/h}$ is valid -- at least theoretically. A detector
consists of three individual detection devices, one for each lane. By
combining three devices covering the three lanes belonging to one
direction (except D2) one gets the cross-sections. D1 and D4 are
installed nearby the highway intersection ``K\"oln-Nord'', while the
others are located close to a junction. These locations are
approximately $9\,\mbox{km}$ apart.  Within two weeks in June 1996
more than two million single-vehicle measurements were recorded.
Nearly $16\,\%$ of the detected vehicles were trucks.
During this period the traffic data set was not biased due to road
constructions or bad weather conditions. The traffic computer recorded
the distance-headway $\Delta x$ as well as the velocity $v$ of the
vehicles passing a detector. The time elapsed between two consecutive
vehicles has to be calculated via $\Delta x_n=v_{n-1}\Delta t_n$ for
the vehicle $n$ and its predecessor $n-1$\footnote{In fact the
  velocity of a car passing the detector and the time elapsed since
  the leading vehicle has passed are the direct measurements (see
  \cite{Neubert} for an elaborate discussion).}.\\
For a sensible discussion it is plausible to split up the data set
according to the different traffic states. Time-series of the speeds
allow for a manual separation into free flow and congested states. The
congested states, in turn, are composed of stop-and-go traffic as well
as of synchronized states (here only type (iii)). A distinction is
possible by utilizing the crosscorrelation as
described later.
Beside the separation with regard to the traffic states the data can
be arranged in density intervals. The density $\rho$ is a global
quantity like flow $J$ or mean velocity $\langle v\rangle$. In the
present case the density can only be determined by the hydrodynamic
relation $\rho=J/\langle v\rangle$. Due to the
event-driven measurements one makes an error in the high-density
regime. Standing cars lead to an overestimation of the velocity of
passing cars and an underestimation of the density. Therefore, the
data points in the fundamental diagram $J(\rho)$
(Fig.\,\ref{fig:autocorr_states}) are shifted to the left in
stop-and-go traffic. In the same way other methods must fail, since
the prime reason is the event-driven
measurement method.\\
Correlation functions can detect and quantify coupling effects between
variables. With $\psi$ the variable of interest the autocorrelation
\begin{eqnarray}
  \label{eq:auto}
  ac_\psi(\tau)=\frac{1}{\sigma(\psi)} \left[ \langle
  \psi(t)\psi(t+\tau)\rangle-\langle\psi(t) \rangle^2 \right]
\end{eqnarray}
provides information about the temporal evolution of $\psi$
($\sigma(\psi)$ is the variance of $\psi$). In
Fig.\,\ref{fig:autocorr_states} the typical behavior for
1-min-averages is depicted. In free-flow traffic there are only slight
changes in both flow and density. These correlations break down during
the crossover to synchronized flow -- the corresponding data points in
the fundamental diagram are scattered over an extended plane. In
stop-and-go traffic an oscillating structure can be found, a hint on
the alternating occurrence of stall and movement \cite{Kuehne}.\\
\begin{figure}
  \begin{center}
    \includegraphics[width=.38\textwidth]{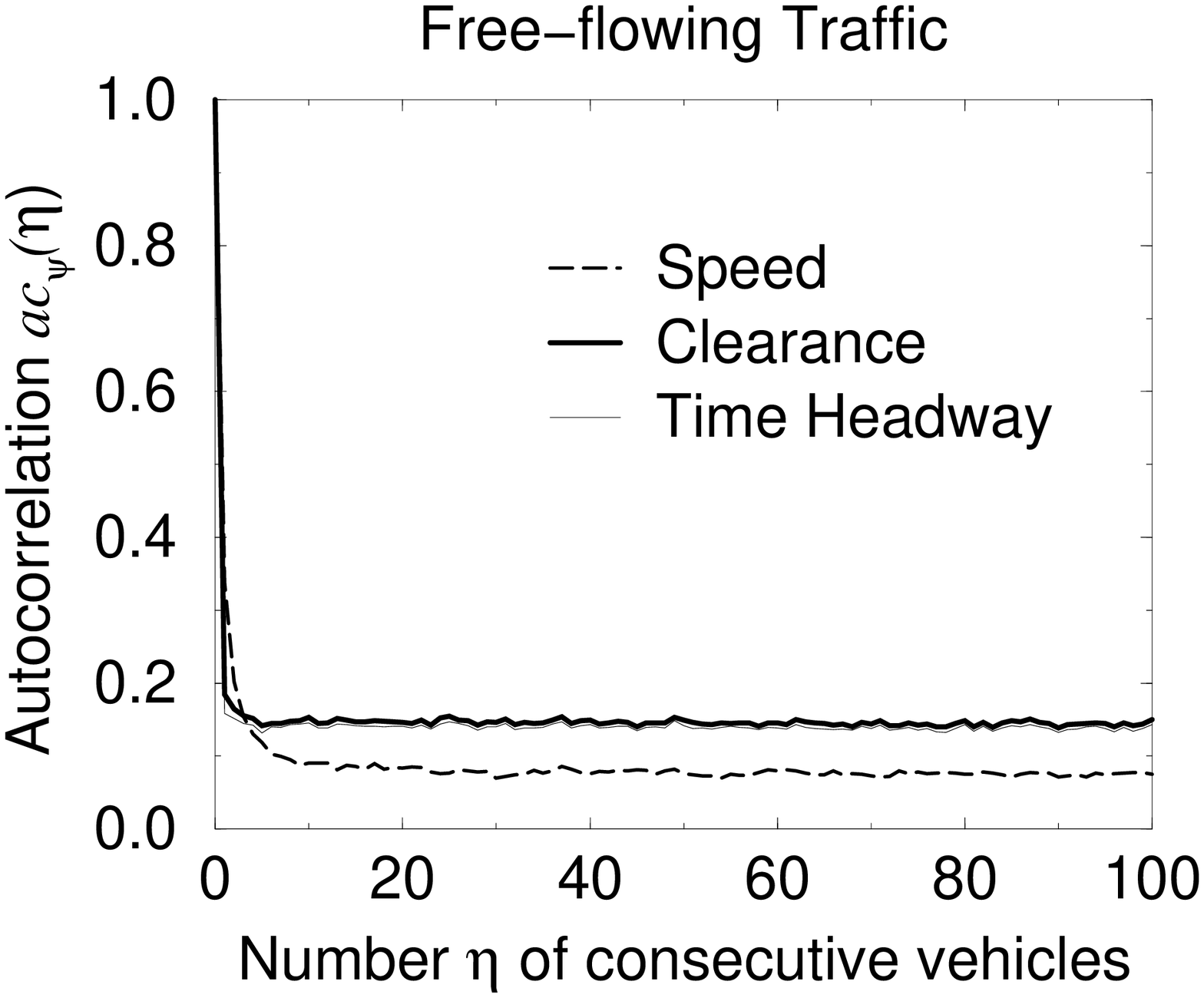}
    \includegraphics[width=.38\textwidth]{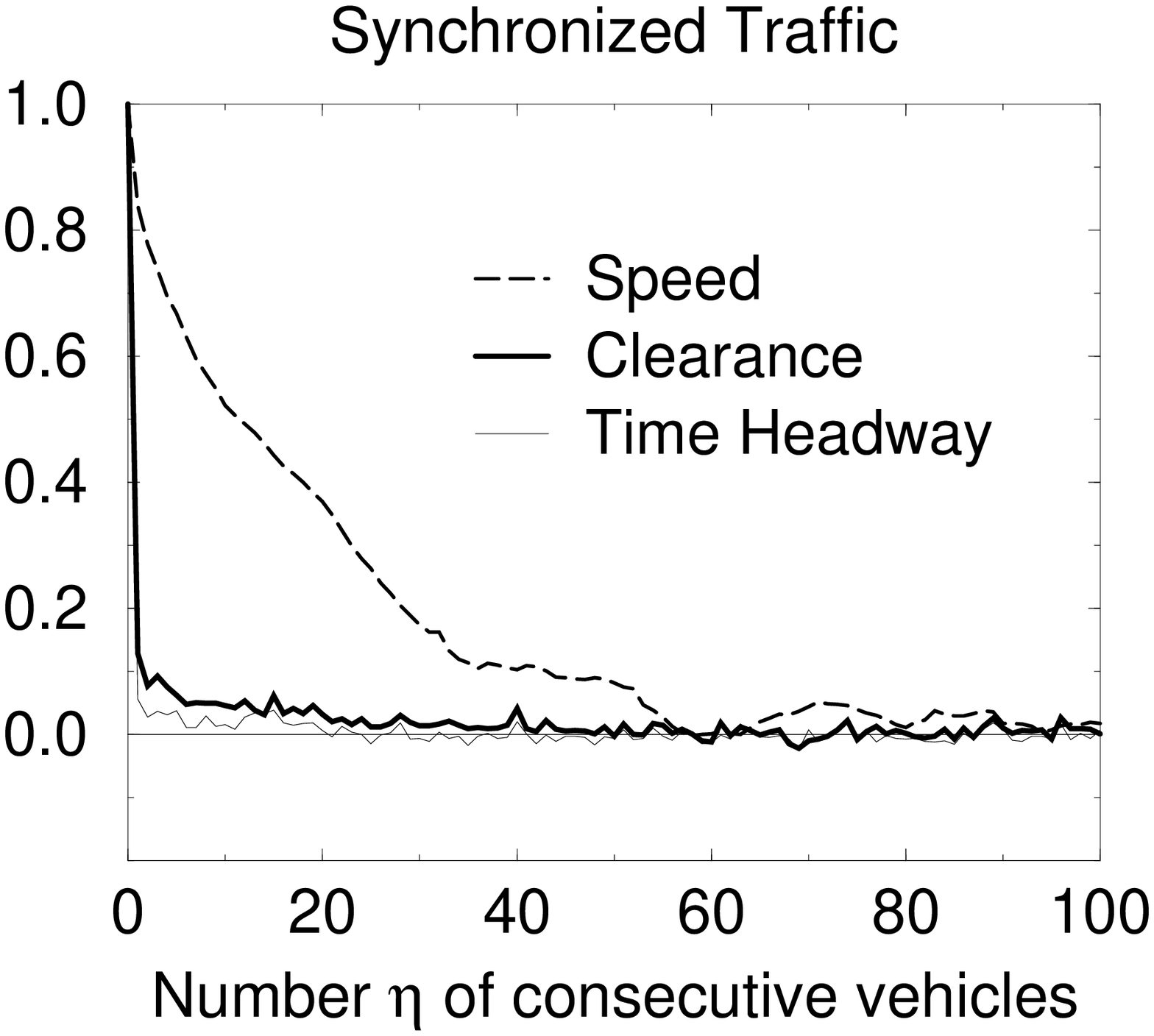}
    \caption{If single-vehicle data are applied to the autocorrelation
      function, one can obtain a synchronization effect along the
      road.}
    \label{fig:autocorr_single}
  \end{center}
\end{figure}
Now, $\tau$ is substituted with $\eta$ in (\ref{eq:auto}) as well as
$t$ with $n$, where $\eta$ is the number of consecutive vehicles. The
results are plotted in Fig.\,\ref{fig:autocorr_single}. In the case of
free-flowing traffic streams the vehicles move quite independent of
each other, which supports the assumption of {\em random headway
  states} for the analytical description of time-headways. If
synchronized traffic emerges an adaption of vehicular speeds between
lanes \cite{Kerner} as well as among consecutive cars is to notice.\\
On the other hand, the crosscorrelation detects, whether two
variables $\xi$ and $\psi$ or coupled or not. It is defined as
\begin{eqnarray}
  \label{eq:cross}
  cc_{\xi\psi}(\tau)=\frac{1}{\sqrt{\sigma(\xi)\sigma(\psi)}}\left[
  \langle \xi(t)\psi(t+\tau)\rangle-\langle\xi(t)\rangle
  \langle\psi(t+\tau)\rangle \right].
\end{eqnarray}
By calculating the crosscorrelation $cc_{J\rho}$ between flow and
density an objective parameter is given for discriminating between the
several states of traffic. From investigations of the present data
base one may conclude the following:
\begin{table}[htbp]
  \vspace*{-0.5cm}
  \begin{center}
    \begin{tabular}{rcl}
      \hline\noalign{\smallskip}
      Traffic state~~~ & $cc_{J\rho}$ & ~~~Interpretation and remarks \\
      \hline\noalign{\smallskip}
      Free flow~~~ & $\approx 1$ & ~~~Density controls flow \\
      Synchronized~~~ & $\approx 0$ & ~~~Data points cover a plane in the
      fundamental diagram \\
      Stop-and-Go~~~ & $\approx\pm 1$ & ~~~Data points close to a line with
      positive or negative slope\\
      & & ~~~(with regard to the applied $\rho$-method)\\
      \hline\noalign{\smallskip}
    \end{tabular}
  \end{center}
\end{table}
Further insights into the car-car-interaction are provided by analyses
of histograms of clearances, speed differences and time headways. Time
headways of unbounded moving cars (Fig.\,\ref{fig:timeheadways}) can
be described through so-called {\em random headway states} which
yields an exponentially decaying background signal of the probability
distribution $P(\Delta t)$ \cite{May} in the form
\begin{eqnarray}
  \label{eq:expo}
  P(\Delta t)=\frac{1}{\tau'}e^{-{\Delta t}/\tau},
\end{eqnarray}
where $\tau$ is directly linked with the present traffic flow $J$ and
$\tau'\approx\tau/\nu$ with $\nu$ the number of bins in an interval of
$1\,\sec$ (here typically $\nu=10$).  Additionally, two peaks show up,
the weaker one near $\Delta t=2\,\sec$ and the higher one below
$1\,\sec$ and rising with ascending density. The latter corresponds to
high-current states in the fundamental diagram, since small
time-headways are necessary to generate them. With the transition to
congested traffic states this background looses its evidence. In
stop-and-go states there is only the peak at $\Delta t \approx
2\,\sec$ -- it reflects the fact that only cars leaving a jam are to
detect with the
typical time headway for this procedure.\\
The data are also suitable to generate {\em OV} diagrams. Before doing
so, it is worth to have a look at the speed differences $\Delta v_n$.
By means of Fig.\,\ref{fig:ov} one soon realizes that the drivers tend
to minimize driving actions and simultaneously to improve their
``driving comfort'': With vanishing $\Delta v_n$ they follow their
predecessors closer than suggested with a mean gap $\propto\rho^{-1}$,
and this is the most probable situation. For deriving an {\em optimal}
velocity only data points with $\Delta v_n\approx 0\,\mbox{m/sec}$ are
taken into account (right diagram in Fig.\,\ref{fig:ov}).
Additionally, the data sets were again
distinguished according to the underlying global traffic state. This
leads to different {\em OV} relations, which can be fitted nicely by a
{\tt tanh} ansatz.

\section{{\em Floating-Car} Data}

\begin{figure}
  \begin{center}
    \includegraphics[width=.38\textwidth]{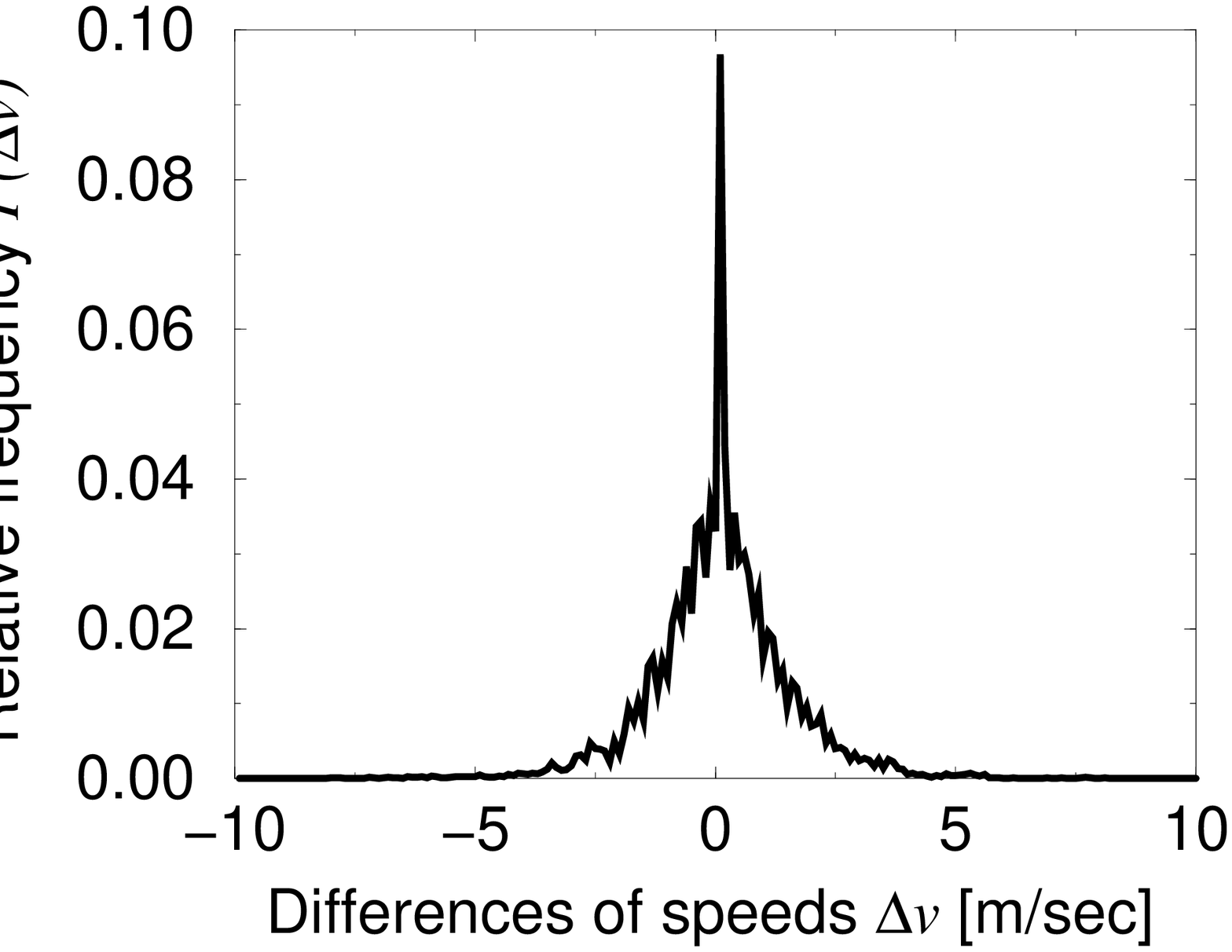}
    \includegraphics[width=.38\textwidth]{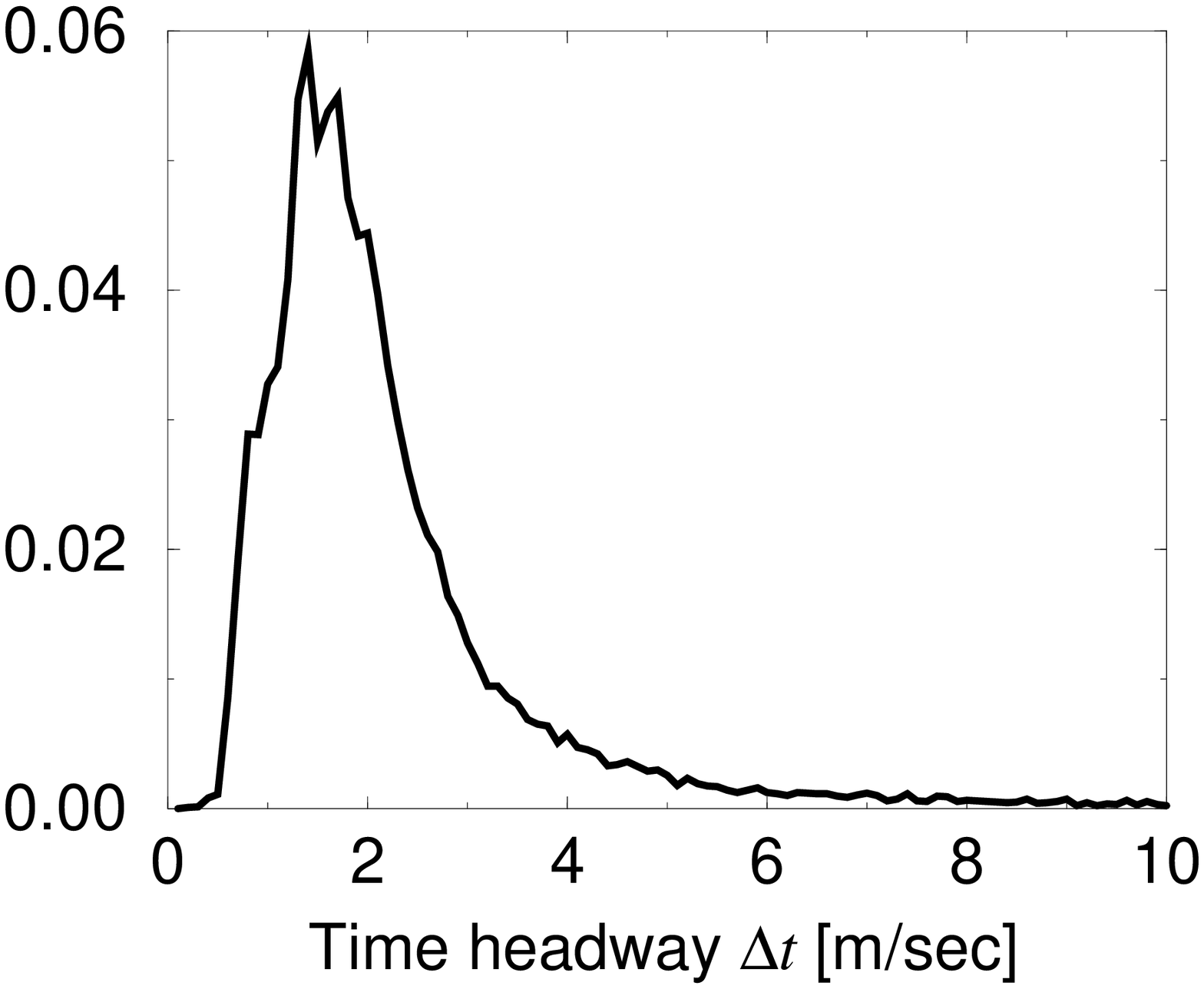}
    \caption{By {\em FC} data previous results can be
      verified. The driver tends to make $\Delta v$ disappear, his
      preferred temporal headway is around $2\,\sec$.}
    \label{fig:ika_histo}
  \end{center}
\end{figure}

The {\em FC} data were collected during field studies
in the framework of the Research-Cooperative ``NRW-FVU'' of
North-Rhine Westfalia \cite{fvu}. The data sets recorded during such
studies contain the own velocity $v$, the difference of velocities
$\Delta v$ to the predecessor and the clearance $\Delta x$ to him. 
\begin{figure}
  \begin{center}
    \includegraphics[width=.38\textwidth]{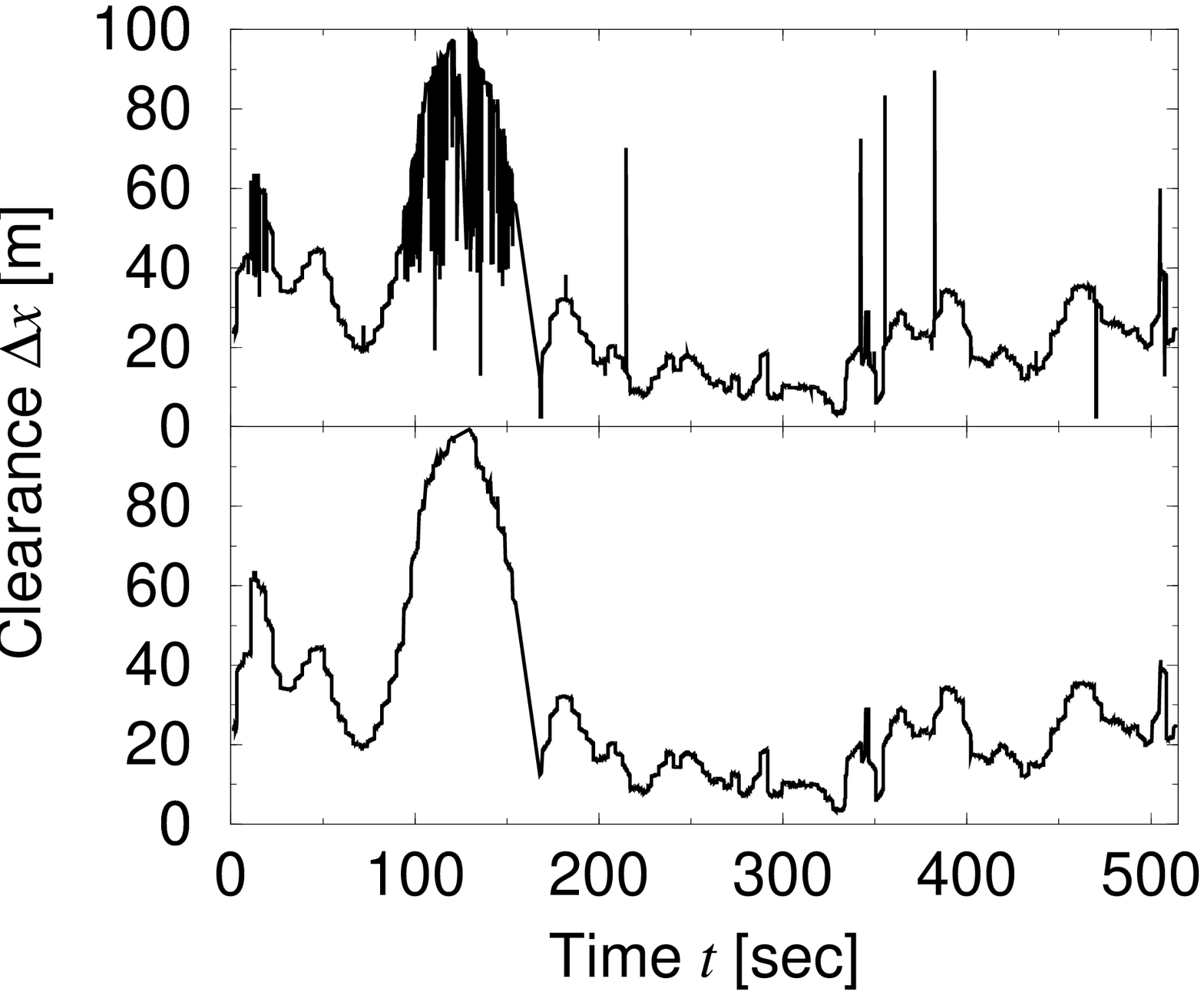}
    \includegraphics[width=.38\textwidth]{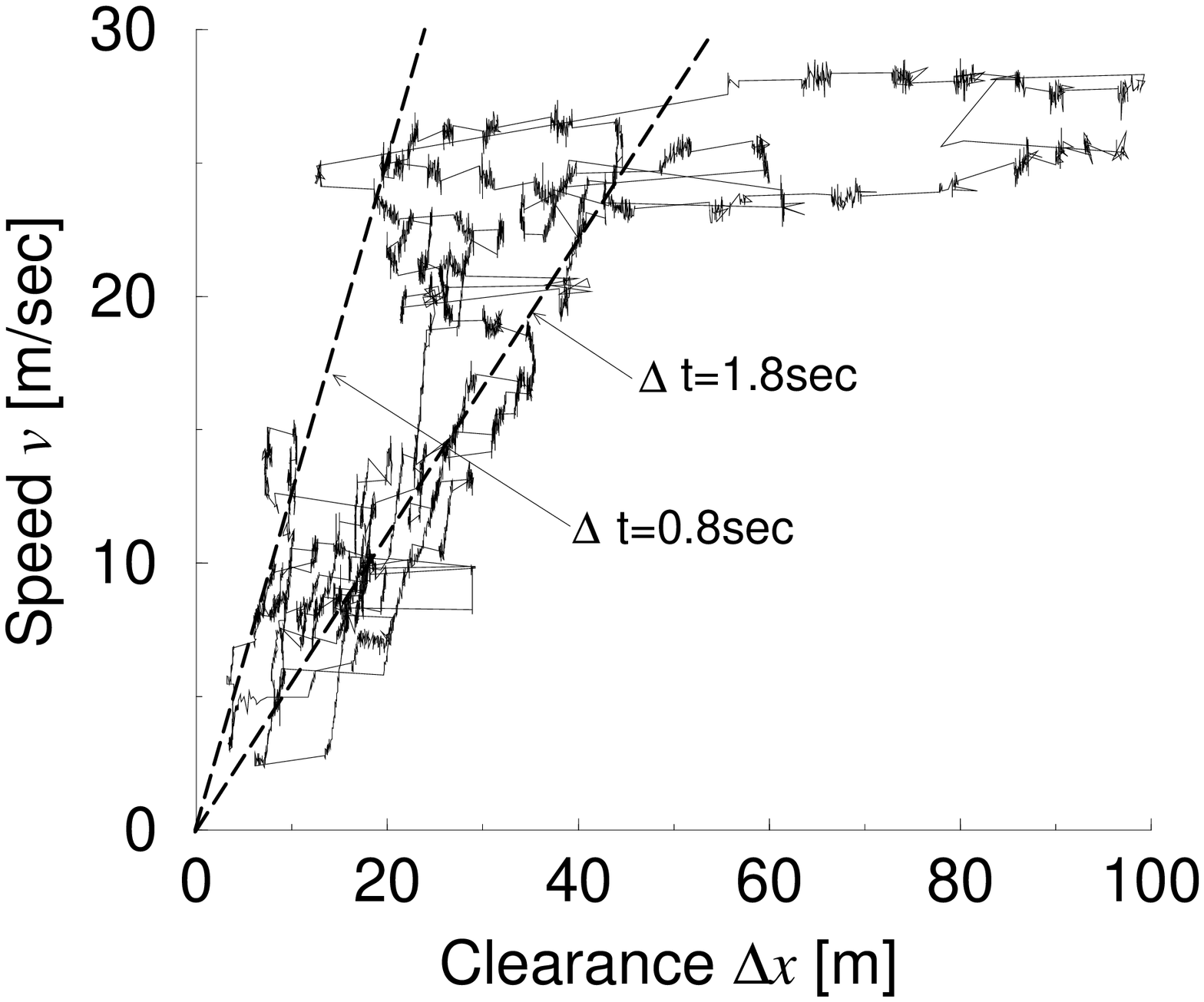}
    \caption{After a clean-up all errors in the time-series were
      erased. With this data base the speed-distance relation (right)
      is plotted.}
    \label{fig:timeseries}
  \end{center}
\end{figure}
The distance to the leading vehicle and its velocity is measured using
a laser beam. This method leads to very precise results as long as the
beam does not miss the leading car. But nevertheless a part of the
data set is not correct, e.g. data recorded during lane-changing
maneuvers. These faulty items can be easily identified and have to be
omitted before analyzing the data set (see
Fig.\,\ref{fig:timeseries}).  Afterwards the data set is diluted in a
way that only one measurement per second remains. The $v(\Delta
x)$-diagram (here without discrimination due to traffic states) now is
comparable to the previously presented one. In free flow the data
points are arranged on a line parallel to the ordinate, very small
clearances (down to $\Delta x\approx 15\,\mbox{m}$ at high speeds
($100\,\mbox{km/h}$)) can be obtained. From this microscopic level of
view also macroscopic quantities can be computed. Since
$J\propto\Delta t^{-1}$ and $\rho\propto\Delta x^{-1}$
\begin{figure}
  \begin{center}
    \includegraphics[width=.38\textwidth]{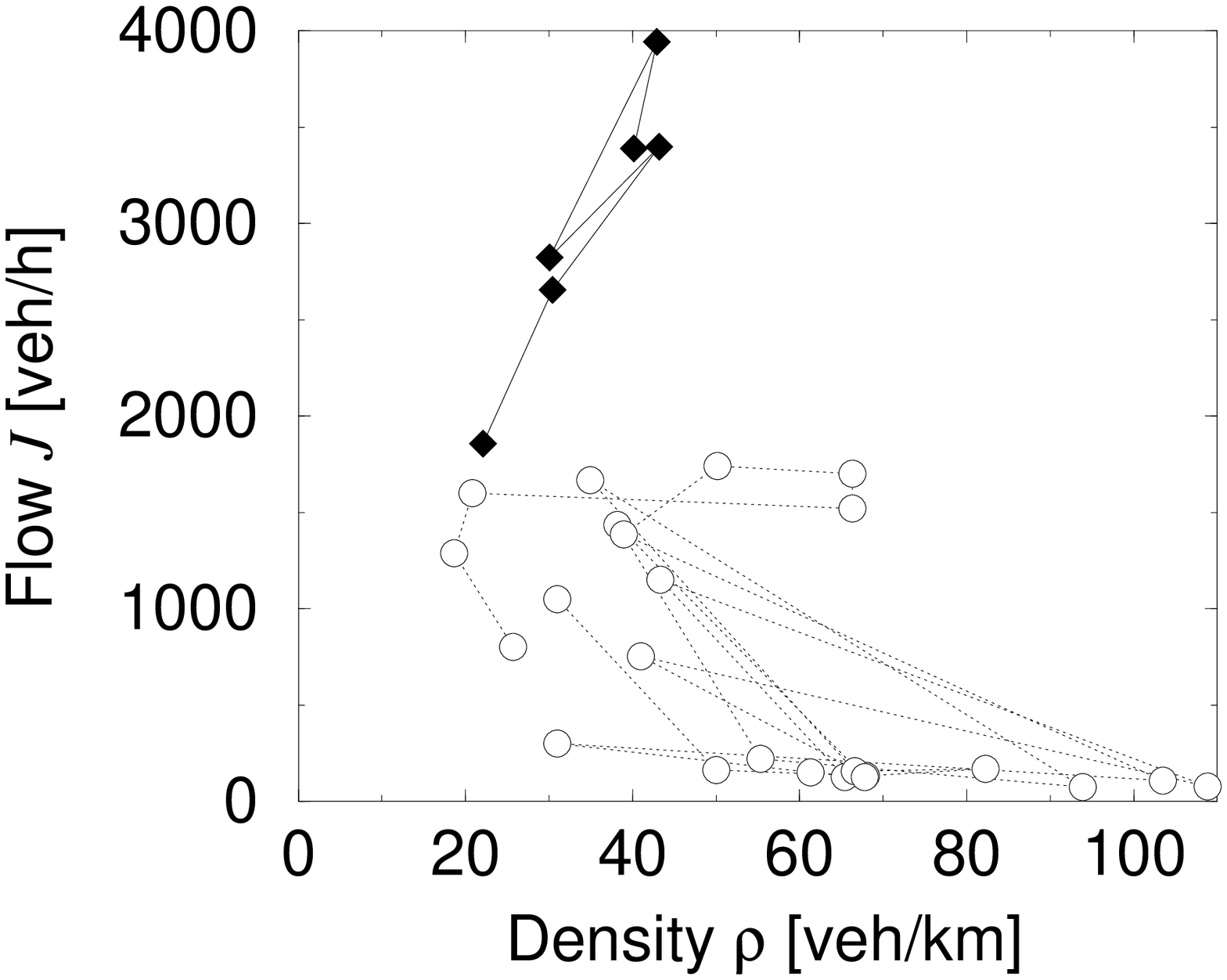}
    \includegraphics[width=.38\textwidth]{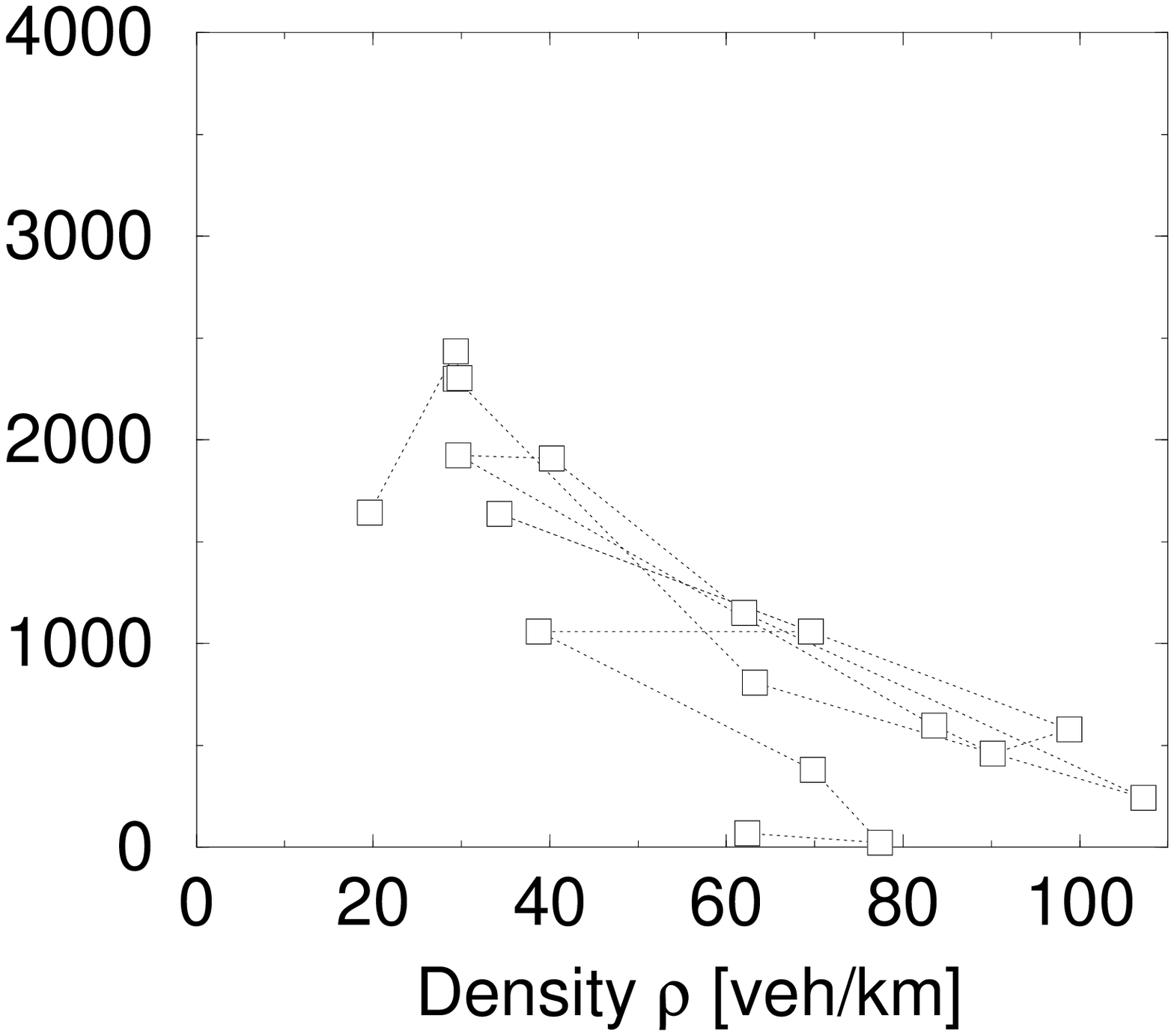}
    \caption{Fundamental diagrams drawn from local measurements. Of
      special interest is the long-lasting high-flow state (left),
      which corresponds to a stable follow-the-leader formation with
      short time-headways for minutes.}
    \label{fig:ika_fd}
  \end{center}
\end{figure}
a fundamental diagram can be plotted (Fig.\,\ref{fig:ika_fd}). Two
important facts have to be mentioned: There are stable high-flow
states persisting for minutes and caused by two consecutive vehicles
with less or no fluctuations in the drivers' behavior. These results
support the view that already in the free-flow regime small platoons
can be observed which move at small distances. The transition into
congested states takes place only at reduced traffic volumes and not
at or nearby the maximum.\\
The histograms in Fig.\,\ref{fig:ika_histo} confirm the findings of
the previous section.  Especially, one recognizes that the state
$\Delta v\approx 0\,\mbox{m/sec}$ is the most attractive one, and the
most probable time headway is located near $\Delta t=2\,\sec$. Crucial
contributions to the distribution come also from time headway shorter
than $1\,\sec$.

\section*{Acknowledgment}
It is our pleasure to thank B.S. Kerner and D. Chowdhury for
fruitful discussions. The authors are grateful to the
``Landschaftsverband Rheinland'' (K\"oln) for data support, to
the ``Systemberatung Povse'' (Herzogenrath) for technical assistance,
to the Ministry of Economic Affairs, Technology and Transport of
North-Rhine Westfalia as well as to the Federal Ministry of Education
and Research of Germany for the financial support (the latter within
the BMBF project ``SANDY''). Parts of this work were done in the
framework of the research-cooperative ``NRW-FVU'' of North-Rhine
Westfalia.

\end{document}